\documentclass{aa}
\usepackage{subfigure}
\usepackage[varg]{txfonts}
\usepackage{natbib,twoopt}
\usepackage[breaklinks=true]{hyperref}
\usepackage{graphicx}
\usepackage{amsmath}
\usepackage{amssymb}
\usepackage{textcomp}
\bibpunct{(}{)}{;}{a}{}{,}
\makeatletter
  \newcommandtwoopt{\citeads}[3][][]{\href{http://adsabs.harvard.edu/abs/#3}%
    {\def\hyper@linkstart##1##2{}%
     \let\hyper@linkend\@empty\citealp[#1][#2]{#3}}}
  \newcommandtwoopt{\citepads}[3][][]{\href{http://adsabs.harvard.edu/abs/#3}%
    {\def\hyper@linkstart##1##2{}%
     \let\hyper@linkend\@empty\citep[#1][#2]{#3}}}
  \newcommandtwoopt{\citetads}[3][][]{\href{http://adsabs.harvard.edu/abs/#3}%
    {\def\hyper@linkstart##1##2{}%
     \let\hyper@linkend\@empty\citet[#1][#2]{#3}}}
  \newcommandtwoopt{\citeyearads}[3][][]%
    {\href{http://adsabs.harvard.edu/abs/#3}
    {\def\hyper@linkstart##1##2{}%
     \let\hyper@linkend\@empty\citeyear[#1][#2]{#3}}}
\makeatother

\def\apm{{APM 08279$+$5255}}
\def\oiii{{[O {\scriptsize III}]}}
\def\h0{{$H_0 = 70$ km s$^{-1}$ Mpc$^{-1}$}}
\def\omegam{{$\Omega_{\rm M} = 0.3$}}
\def\omegalambda{{$\Omega_\Lambda = 0.7$}}
\def\si4{{Si {\scriptsize IV}}}
\def\civ{{C {\scriptsize IV}}}
\def\c3{{C {\scriptsize III}]}}
\def\al3{{Al {\scriptsize III}}}
\def\mg2{{Mg {\scriptsize II}}}
\def\hbeta{{H$\beta$}}
\def\fe2{{Fe {\scriptsize II}}}

\title{The rest-frame UV-to-optical spectroscopy of \apm}
\subtitle{BAL classification and black hole mass estimates}

\author{{F.~G. Saturni}\inst{\ref{inst1},\ref{inst2}}
\and
{M. Bischetti}\inst{\ref{inst1},\ref{inst3}}
\and
{E. Piconcelli}\inst{\ref{inst1}}
\and
{A. Bongiorno}\inst{\ref{inst1}}
\and
{C. Cicone}\inst{\ref{inst4}}
\and
{C. Feruglio}\inst{\ref{inst5}}
\and
{F. Fiore}\inst{\ref{inst1}}
\and
{S. Gallerani}\inst{\ref{inst6}}
\and\\
{M. Giustini}\inst{\ref{inst7}}
\and
{S. Piranomonte}\inst{\ref{inst1}}
\and
{G. Vietri}\inst{\ref{inst1},\ref{inst8},\ref{inst9}}
\and
{C. Vignali}\inst{\ref{inst10},\ref{inst11}}
}
\institute{INAF -- Osservatorio Astronomico di Roma, Via Frascati 33, I-00040 Monte Porzio Catone (RM), Italy.\\
\email{francescogabriele.saturni@oa-roma.inaf.it}\label{inst1}
\and
Space Science Data Center, Agenzia Spaziale Italiana, Via del Politecnico snc, I-00133 Roma, Italy.\label{inst2}
\and
Dip. di Fisica, Universit{\`a} degli Studi di Roma ``Tor Vergata'', Via della Ricerca Scientifica 1, I-00133 Roma, Italy.\label{inst3}
\and
INAF -- Osservatorio Astronomico di Brera, Via Brera 28, I-20121 Milano, Italy.\label{inst4}
\and
INAF -- Osservatorio Astronomico di Trieste, Via G. B. Tiepolo 11, I-34143 Trieste, Italy.\label{inst5}
\and
Scuola Normale Superiore, P.zza dei Cavalieri 7, I-56126 Pisa, Italy.\label{inst6}
\and
SRON -- Netherlands Institute for Space Research, Sorbonnelaan 2, NL-3584 CA Utrecht, The Netherlands.\label{inst7}
\and
Excellence Cluster {\itshape Universe}, Technische Universit{\"a}t M{\"u}nchen, Boltzmannstr. 2, D-85748 Garching b. M{\"u}nchen, Germany.\label{inst8}
\and
European Southern Observatory, Karl-Schwarzschild-Str. 2, D-85748 Garching b. M{\"u}nchen, Germany.\label{inst9}
\and
Dip. di Fisica e Astronomia, {\itshape Alma Mater Studiorum}, Universit{\`a} degli Studi di Bologna, Via P. Gobetti 93/2, I-40129 Bologna, Italy.\label{inst10}
\and
INAF -- Osservatorio di Astrofisica e Scienza dello Spazio di Bologna, Via P. Gobetti 93/3. I-40129 Bologna, Italy.\label{inst11}
}
\date{Received 2018 Feb 08 / Accepted 2018 Apr 24}

\abstract
	{We present the analysis of the rest-frame optical-to-UV spectrum of \apm, a well-known lensed broad absorption line (BAL) quasar at $z = 3.911$. The spectroscopic data are taken with the optical DOLoRes and near-IR NICS instruments at {\itshape TNG}, and include the previously unexplored range between \c3\ $\lambda$1910 and \oiii\ $\lambda\lambda$4959,5007. We investigate the possible presence of multiple BALs by computing ``balnicity'' and absorption indexes (i.e. BI, BI$_0$ and AI) for the transitions \si4\ $\lambda$1400, \civ\ $\lambda$1549, \al3\ $\lambda$1860 and \mg2\ $\lambda$2800. No clear evidence for the presence of absorption features is found in addition to the already known, prominent BAL associated to \civ, which supports a high-ionization BAL classification for \apm. We also study the properties of the \oiii, \hbeta\ and \mg2\ emission lines. We find that \oiii\ is intrinsically weak ($F_{\rm [OIII]}/F_{\rm H\beta} \lesssim 0.04$), as it is typically found in luminous quasars with a strongly blueshifted \civ\ emission line ($\sim$2500 km s$^{-1}$ for \apm). We compute the single-epoch black hole mass based on \mg2\ and \hbeta\ broad emission lines, finding $M_{\rm BH} = (2 \div 3) \times 10^{10}\mu^{-1}$ M$_\odot$, with the magnification factor $\mu$ that can vary between 4 and 100 according to CO and rest-frame UV-to-mid-IR imaging respectively. Using a \hbox{\mg2} equivalent width (EW)-to-Eddington ratio relation, the EW$_{\rm MgII} \sim 27$ \AA~measured for \apm\ translates into an Eddington ratio of $\sim$0.4, which is more consistent with $\mu=4$. This magnification factor also provides a value of $M_{\rm BH}$ that is consistent with recent reverberation-mapping measurements derived from \civ\ and \si4.}


\begin{document}
\titlerunning{Rest-frame UV-to-optical spectroscopy of APM 08279+5255}
\authorrunning{F.~G. Saturni et al.}
\maketitle

\section{Introduction}\label{intro}

\apm\ is a well-known luminous broad absorption-line quasar (BAL QSO) at $z = 3.911$. Serendipitously discovered in a Galactic survey for cold carbon stars \citep{Irw98}, it is archetypal to several categories of the quasar class, showing together many of the observational phenomena that can be found in such objects. In fact, beyond having evidence of both broad \citep{Sri00} and intrinsic narrow \citep{Ell04} absorption features associated with the \civ\ $\lambda$1549 emission line, it also shows an uncommon O {\scriptsize VI} $\lambda$1030 BAL embedded in the Ly$\alpha$ forest \citep{Hin99} and an X-ray ultra-fast outflow (UFO) associated to highly-ionized iron \citep{Has02,Cha02,Sae09,Hag17}. Furthermore, its high-ionization emission lines, from Ly$\alpha$ to \c3\ $\lambda$1910, are characterized by a significant blueshift of $\sim$2500 km s$^{-1}$ with respect to molecular \citep{Dow99} and Balmer lines \citep{Oya09}. A blueshifted emission component with $v \sim -800$ km s$^{-1}$, corresponding to a molecular outflow, is also detected in the CO($4-3$) transition by \citet{Fer17} through 3.2 mm observations with the NOEMA interferometer.

\begin{figure*}[htbp]
\begin{center}
\includegraphics[scale=0.65,angle=-90]{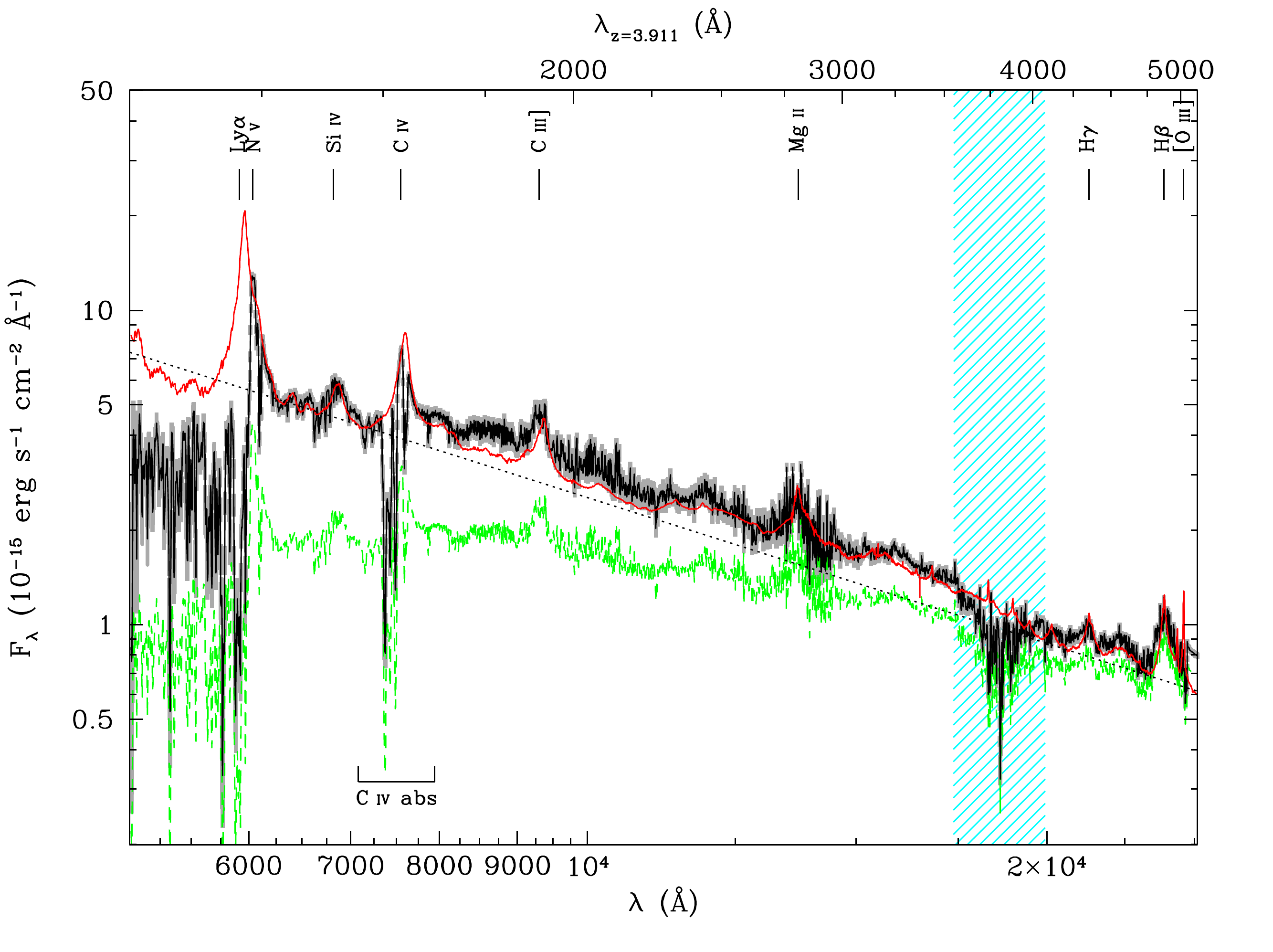}
\end{center}
\caption{\apm\ calibrated spectrum before ({\itshape green dashed line}) and after dereddening ({\itshape black solid line}), with corresponding 1$\sigma$ uncertainty ({\itshape grey band}) -- see Sec. \ref{obsred}. Superimposed to the data, the composite quasar template obtained by matching the \citet{Van01} and \citet{Gli06} templates around $\lambda \sim 3000$ \AA~in the rest frame ({\itshape red line}) is shown. The continuum level is marked as a black dotted line with slope $-1.54$. A residual telluric absorption is still present between $\lambda \sim 17,500$ \AA~and $20,000$ \AA~({\itshape cyan shaded area}), hence this spectral interval is excluded from the present analysis.}
\label{totspec}
\end{figure*}

\apm\ is gravitationally lensed \citep{Led98} by an unobserved galaxy at $z \sim 1$ \citep{Pet00,Ell04}. The lensed image is elongated in the NE direction and consists of three components, with a maximum separation of $0''.35 \pm 0''.02$ \citep{Led98}. This makes \apm\ the first confirmed case with an odd number of images \citep{Iba99,Lew02a}. The lack of knowledge about the lensing object further complicates the estimation of the magnification factor $\mu$. Current lens models are built on the observation of the CO($1-0$) molecular line \citep[e.g.,][]{Lew02b}, but the resulting $\mu$ is strongly dependent on the lens geometry, ranging from $\mu \sim 4$ \citep[highly-inclined spiral galaxy;][]{Rie09} up to $\mu \sim 100$ \citep[naked cusp;][]{Ega00}. Nevertheless, with an intrinsic bolometric luminosity in the range of $5 \times 10^{13}$ to $10^{15}$ L$_\odot$ \apm\ can be considered among the most intrinsically luminous quasars.

Since its discovery, several photometric and spectroscopic observational campaigns have targeted \apm\ in different energy bands. Both short-term and long-term monitoring of this object have been performed for a wide range of purposes, from the analysis of its optical variability \citep{Lew99} to the study of the UFO through photoionization codes \citep{Sae11}, the investigation of the \civ\ absorption variability \citep{Tre13,Sat14} and the reverberation mapping \citep{Tre07,Sat16}. In particular, the variability study of \apm\ absorption systems \citep{Tre13,Sat14,Sat16} concluded that the \civ\ absorption variability in \apm\ is most likely driven by changes in the photoionization state of the gas, responding to variations of the \civ\ ionizing continuum level.

Single-epoch observations of \apm\ include rest-frame UV high-resolution spectroscopy with {\itshape Keck}/HIRES \citep{Ell99} and {\itshape HST}/STIS \citep{Lew02a} for the study of the damped Ly$\alpha$ absorbers (DLAs) and intervening absorption systems \citep{Pet00,Ell04}. \citet{Sri00} first analyzed the high-velocity absorption system bluewards the \civ\ emission peak thanks to the availability of the {\itshape Keck}/HIRES spectrum, finding narrow absorption lines embedded between two unresolved broad components. Using the same spectrum, \citet{Ell04} studied the resolved absorption feature on the \civ\ red wing, classifying it as a system of four intervening clouds located close to the quasar systemic redshift.

In this paper, we present the quasi-simultaneous rest-frame optical-to-ultraviolet (UV) spectrum of \apm\ taken at the 3.5 m {\itshape Telescopio Nazionale Galileo} ({\itshape TNG}) in La Palma (Canarian Islands) with the Device Optimized for Low Resolution (DoLoRES; $\lambda/\Delta\lambda \sim 700$) and the Near-Infrared Spectrograph and Camera (NICS; $\lambda/\Delta\lambda \sim 500$). Covering the region between \c3\ and \oiii\ $\lambda\lambda$4959,5007 which was unobserved so far, this broad-band spectrum allows the study of the rest-frame wavelength range \hbox{$\lambda\lambda \sim$ 1000 -- 5000 \AA} in a single state of quasar activity. In fact, the interval of 76 days between the near-infrared (NIR) and optical observations corresponds to a rest-frame interval of $\sim$15 days, much shorter than typical variability timescales of \apm\ \citep[$\sim$430 rest-frame days for continuum flux changes; e.g.,][]{Sat16}. The paper is organized as follows: we describe the observations and the procedure of data reduction in Sec. \ref{obsred}; we analyze the spectral features in Sec. \ref{niropt}; finally, we present single-epoch supermassive black hole (SMBH) mass estimates in Sec. \ref{resconc} and discuss our results in Sec. \ref{disc}. Throughout the text, we report all errors at $1\sigma$ confidence level, and adopt a concordance cosmology with \hbox{\h0,} \omegam\ and \omegalambda.

\section{Observations and data reduction}\label{obsred}

Observations of the \apm\ rest-frame optical-to-UV spectrum were carried out on 2011 February 19-20 (optical) and 2011 May 05 (UV) at {\itshape TNG}. The rest-frame UV spectrum of \apm\ was obtained with the $R$-band grism (wavelength range $\lambda\lambda$4470 -- 10,073 \AA, dispersion of 2.61 \AA~px$^{-1}$, $\lambda / \Delta \lambda = 714$) of the DOLoRes instrument, coupled to the 1$''$ slit. The rest-frame optical spectrum was acquired with the NICS instrument in two low-resolution configurations, respectively for the $IJ$ (wavelength range $\lambda\lambda$9000 -- 14,500 \AA, dispersion of \hbox{5.5 \AA~px$^{-1}$,} $\lambda / \Delta \lambda = 500$) and $HK$ (wavelength range \hbox{$\lambda\lambda$14,000 -- 25,000 \AA,} dispersion of 11.2 \AA~px$^{-1}$, $\lambda / \Delta \lambda = 500$) bands, with the same slit width of the $R$-band spectrum.

The rest-frame UV spectrum $\lambda\lambda 1020 - 1870$ \AA\ considered in our analysis extends over the observed wavelength interval $\lambda\lambda \sim 5000 - 9200$ \AA. A large contamination due to overlap of higher spectral orders is visible redwards $\lambda \sim 9200$ \AA. The spectrum was calibrated with standard {\scriptsize IRAF} procedures, and was cleaned from the major telluric absorptions, namely the Fraunhofer A and B bands and the H$_2$O features, adopting the method described in \citet{Tre13}. The total rest-frame optical spectrum extends over the observed wavelength intervals \hbox{$\lambda\lambda \sim 8700 - 14,500$ \AA} ($IJ$ bands) and $\lambda\lambda \sim 13,500 - 24,700$ \AA~($HK$ bands), corresponding to $\lambda\lambda$1770,5030 \AA~in the rest-frame UV-to-optical bands. A standard NICS observing sequence consists of exposures at two different dither positions ($A$ and $B$) and taken in the pattern $ABBA$. Background subtraction was obtained performing $A-B$ and $B-A$ image differences, obtaining four positive aperture images. The spectrum was extracted for each differential image in a standard way using the {\scriptsize IRAF} task {\ttfamily apall}. Then, in order to remove cosmic rays, the four extracted spectra were combined together. Finally, a telluric standard star was used to correct the target spectrum for the atmospheric transmission.

\begin{table}[htbp]
\begin{center}
\begin{tabular}{lcccc}
\hline
\hline
\multicolumn{5}{l}{}\\
Band & $\lambda_{\rm eff}$ (\AA) & $m_{\rm ph}$ (mag) & $\Delta m_{\rm ph}$ (mag) & Ref.\\
\multicolumn{5}{l}{}\\
\hline
\multicolumn{5}{l}{}\\
$B$ & 4380 & 18.827 & 0.017 & 1\\
$V$ & 5450 & 16.448 & 0.012 & 1\\
$R$ & 6410 & 15.353 & 0.014 & 1\\
$I$ & 7980 & 14.608 & 0.012 & 1\\
$J$ & 12,200 & 13.340 & 0.030 & 2\\
$H$ & 16,300 & 12.650 & 0.030 & 2\\
$K_{\rm s}$ & 21,300 & 12.080 & 0.030 & 2\\
$L'$ & 34,500 & 9.900 & 0.040 & 2\\
\multicolumn{5}{l}{}\\
\hline
\multicolumn{5}{l}{{\scriptsize $^1$\citet{Ojh09}}}\\
\multicolumn{5}{l}{{\scriptsize $^2$\citet{Ega00}}}\\
\end{tabular}
\end{center}
\caption{Optical-to-near infrared Vega magnitudes of \apm\ in the observer frame available in the literature.}
\label{mags}
\end{table}

In producing the joint \apm\ UV-to-optical spectrum, we compared the rest-frame UV flux with a coeval spectrum taken in April 2011 at the Asiago observatory (Italy) for a reverberation-mapping campaign of luminous quasars \citep{Tre07,Tre14,Sat16}. We noted that the {\itshape TNG} UV flux level obtained from standard-star calibration was a factor $\sim$1.1 lower than the Asiago spectrum, which was acquired together with a reference star within a wide ($8''$) slit in order to do not generate differential light losses and be therefore able to construct meaningful light curves \citep[see e.g.][for a discussion]{Kas07}. We suspected that a light loss happened in acquiring the {\itshape TNG}/DOLoRes spectrum due to a seeing-limited observation rather than diffraction-limited (seeing at La Palma site of up to $\sim$1$''$.5, to be compared with the $1''$ slit width). Additionally, no IR photometric standard stars were available during the NICS observing night. Therefore, we decided instead to recalibrate the \apm\ full spectrum to the photometry reported in \citet{Ega00,Ojh09}.

First, we matched the spectral sections together by scaling them to the integrated fluxes computed in intervals around two fiducial wavelengths, specifically $\lambda \sim 9200$ \AA~for the match between $R$ and $IJ$ bands and $\lambda \sim 14,400$ \AA~for the match between $IJ$ and $HK$ bands. The matching wavelengths were selected by visual inspection, in order to identify overlapping spectral regions relatively free from fringing and superposition of contiguous spectral orders. The flux-calibrated spectrum was finally obtained by normalizing the joint sections to the corresponding photometry of \hbox{\apm} available in the literature (see Tab. \ref{mags}; $BVRI$ magnitudes are from \citealt{Ega00}, $JHK_{\rm s}L'$ magnitudes are from \citealt{Ojh09}). To do so, we produced spectro-photometric points by integrating the joint spectrum over the bands listed in \hbox{Tab. \ref{mags}}; then, we performed a 1st-order spline fit to the ratios between \apm\ literature magnitudes and the spectro-photometric points, and obtained the final spectrum by multiplying the joint spectrum by this spline. This approach does not introduce a significant amount of additional uncertainty on the spectral flux level, since the average photometric error $\Delta m_{\rm ph}$ listed in Tab. \ref{mags} corresponds to a flux error of $\sim 9 \times 10^{-17}$ erg s$^{-1}$ cm$^{-2}$, which is comparable to the average rms spectrum over the whole wavelength range.

We checked that the magnitudes of \apm\ used in the recalibration were measured in epochs in which the quasar continuum is not varying. The near-infrared magnitudes are taken at MJD $= 51,089$ \citep{Ega00}, whereas the optical photometry is measured in runs between MJD $= 53,440$ and 54,124 \citep{Ojh09}. Comparing these epochs with those of the $R$-band photometric observations used to construct the light curve of \apm\ in \citet{Tre13} and \citet[][see their fig. 2]{Sat16}, we note that they fall in periods during which the observer-frame optical flux of the quasar remains constant within 0.04 mag. Therefore, our photometry-based spectral recalibration is free of biases introduced by spectral variability.

In order to remove the dust reddening, the spectrum was further de-reddened according to a Small-Magellanic-Cloud (SMC) extinction law \citep{Pei92} with $A_{\rm V} = 0.6$ \citep{Pet00} at $z = 1.062$, i.e. the probable redshift of the lensing galaxy \citep{Ell04}. Fig. \ref{totspec} shows the comparison of the joint rest-frame optical-to-UV spectrum of \apm\ taken at {\itshape TNG} with a template obtained by matching the \citet{Van01} and the \citet{Gli06} SDSS broad-line quasar templates around $\lambda \sim 3000$ \AA~in the rest frame, and normalized to match the observed flux at 1350 \AA~in the rest frame. In our procedure, we have not accounted for the intrinsic reddening of the host galaxy \citep[e.g.,][]{Gal10}. In general, BAL QSOs appear in fact to be more reddened by intrinsic dust with respect to normal quasars \citep[see e.g.][and refs. therein]{Ric03}, with a BAL QSO fraction rising up to $\sim$40\% in extremely reddened objects \citep[e.g.,][]{Urr09}. However, the comparison of our de-reddened spectrum with the joint quasar template (which is in turn used in the following to compute the indexes of absorption for \apm) provides no evidence for intrinsic dust reddening within the host galaxy. We thus prefer to consider only the reddening from the lensing galaxy, although we cannot rule out a possible contribution from the \apm\ host.

\section{UV and optical spectral feature analysis}\label{niropt}

\subsection{The absorption features}\label{balfts}

The rest-frame UV-to-optical spectrum of \apm\ is suitable to study the presence of broad absorption features other than those associated to \civ. To this purpose, we calculate the indexes of absorption most commonly used to identify and classify BAL QSOs, namely the ``balnicity'' index (BI; \citealt{Wey91}), the zero-velocity ``balnicity'' index  (BI$_0$; \citealt{Gib09}) and the absorption index (AI; \citealt{Tru06}). We perform this calculation for both the high- and low-ionization transitions that most frequently produce absorption in this class of objects, i.e. \si4\ $\lambda$1400, \civ\ $\lambda$1549, \al3\ $\lambda$1860 and \hbox{\mg2\ $\lambda$2800}.

The mathematical expression of the BI, BI$_0$ and AI can be generalized into an integral quantity $I(\mathbf{k})$, where $\mathbf{k} = (k_{1 \rightarrow 5})$ represents a set of parameters that define the integration limits, the minimal depth and the minimal velocity width of the absorption. We call $I(\mathbf{k})$ the generalized index of absorption, and define it as follows:
\begin{equation}\label{indgen}
I\left(
\mathbf{k}
\right) = -\int_{-3000 k_1}^{-(25 + 4 k_2)\times 1000}\left[
1-\frac{f(v)}{0.1 k_3}
\right]C\left(
k_4, k_5
\right) dv\mbox{,}
\end{equation}
where $f(v)$ is the normalized QSO flux in the velocity space, and $C(k_4,k_5)$ a constant assuming unitary value over absorbed regions in which the integrand function is positive, provided that the integrand itself remains greater than $0.1 k_4$ in contiguous portions of the absorption trough at least $1000 k_5$ km s$^{-1}$ wide (otherwise, $C = 0$). Within this scheme, the set $\mathbf{k}$ uniquely identifies each index of absorption: $\mathbf{k}_{\rm BI} = (1,0,9,0,2)$, $\mathbf{k}_{\rm BI_0} = (0,0,9,3,2)$ and $\mathbf{k}_{\rm AI} = (0,1,10,1,1)$. To compute these indexes, we normalize the dereddened \hbox{\apm} spectrum to the joint quasar template from \citet{Van01} and \citet{Gli06} shown in Fig. \ref{totspec}, then evaluating the BI, BI$_0$ and AI according to Eq. \ref{indgen}.

The formal error $\sigma^2_I(\mathbf{k})$ associated to $I(\mathbf{k})$ is connected to the rms error on the flux $\sigma_f(v)$ by:

\begin{table}[htbp]
\begin{center}
\resizebox{\columnwidth}{!}{
\begin{tabular}{lccc}
\hline
\hline
\multicolumn{4}{l}{}\\
Transition & BI (km s$^{-1}$) & BI$_0$ (km s$^{-1}$) & AI (km s$^{-1}$)\\
\multicolumn{4}{l}{}\\
\hline
\multicolumn{4}{l}{}\\
\si4\ & $<$70 &  --- & $<$230\\
\civ\ & $3370 \pm 200$ & $2430 \pm 300$ & $4210 \pm 240$\\
\al3\ & $<$10 & --- & $<$110\\
\mg2\ & $<$170 & --- & $<$490\\
\multicolumn{4}{l}{}\\
\hline
\end{tabular}
}
\end{center}
\caption{\apm\ indexes of absorption associated to the main BAL transitions in quasars. Locations marked with `---' indicate indexes for which a reliable upper limit cannot be provided.}
\label{index}
\end{table}

\begin{equation}\label{inderr}
\sigma^2_I\left(
\mathbf{k}
\right) = -\int_{-3000 k_1}^{-(25 + 4 k_2)\times 1000}\left[
\frac{\sigma_f(v)}{0.1 k_3}
\right]^2 C\left(
k_4, k_5
\right) dv,
\end{equation}
although the real uncertainty is usually dominated by systematics in the continuum placement \citep{Tru06}. Therefore, we adopt a Monte-Carlo simulative approach to give a proper evaluation of the uncertainties associated to each non-zero index, or provide fiducial upper limits. Accordingly, we alter the spectrum by adding random noise with Poissonian distribution to each spectral bin, which is assumed to be the mean value of the noise distribution at its wavelength. We then recompute $I(\mathbf{k})$ on this altered spectrum, iterating the process $10^3$ times to reach statistical significance. Finally, we take the standard deviation of the $I(\mathbf{k})$ posterior distribution as the uncertainty to be associated to a nonzero index on the true spectrum. In case of null indexes, we set this standard deviation as the upper limit on the absorption strength. This procedure succeeds in producing fiducial uncertainties or upper limits for all the indexes of absorption but the BI$_0$ for \si4, \al3\ and \mg2. This can be explained in terms of  the conservative definition of BI$_0$ given by \citet{Gib09}, which requires deep absorption features to produce a positive value of this index.

\begin{figure*}[htbp]
\begin{center}
\includegraphics[scale=0.65,angle=-90]{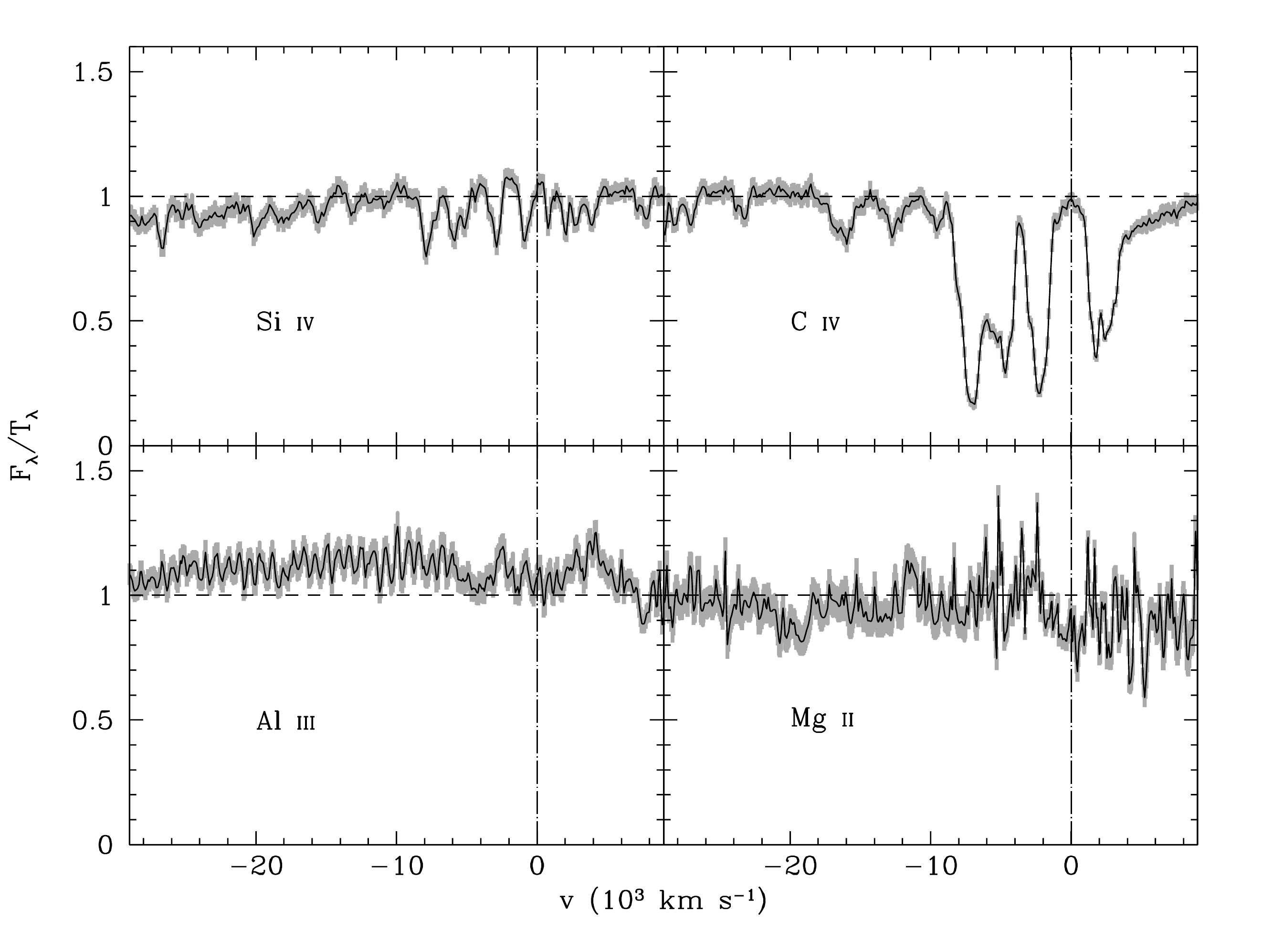}
\end{center}
\caption{Flux of \apm\ normalized to the \citet{Van01} and \citet{Gli06} joint quasar template. The spectral regions around the \si4\ emission ({\itshape top left panel}), the \civ\ emission ({\itshape top right panel}), the \al3\ emission ({\itshape bottom left panel}) and the \mg2\ emission ({\itshape bottom right panel}) are shown. In all panels, the velocity scale is relative to the systemic redshift $z = 3.911$ derived from the CO($4-3$) and CO($9-8$) emission lines \citep{Dow99}. As a guidance, the zero-velocity position ({\itshape dot-dashed line}) and the normalized flux level ({\itshape dashed line}) are indicated.}
\label{normabs}
\end{figure*}

Fig. \ref{normabs} shows the normalized spectral regions around \hbox{\si4,} \hbox{\civ,} \al3\ and \mg2. The values of BI, BI$_0$ and AI for each transition are listed in Tab. \ref{index}, along with the associated uncertainty. We consider significant only absorption detected at a confidence level $>$3$\sigma$. The only absorption feature that satisfies this criterion is the known BAL associated to \civ. However, when observed at high resolution, this absorption structure reveals to be not a single trough, but rather a complex system with two true \civ\ BALs at $v \sim -9750$ and $\sim -4500$ km s$^{-1}$ separated by narrow absorptions located around \hbox{$v \sim -8500$ km s$^{-1}$} \citep{Sri00}. Variability studies suggest that this narrow-absorption complex is nevertheless associated with \civ\ outflows, since it shows a variability pattern very similar to the \hbox{\civ} BAL \citep[see e.g. fig. 6 of][]{Tre13}. In this case the integral in Eq. \ref{indgen} is calculated across the whole spectral range $\lambda\lambda 1400 - 1550$, hence including all the features and giving values in agreement with those obtained by \citet{Tre13} for the equivalent width (EW) of the total absorption (from $\sim$2300 to $\sim$4800 km s$^{-1}$). In addition, the upper limits to the \si4\ absorption strength reflect the sampling of the multiple narrow absorption identified by \citet{Ell04} as intervening \civ, \si4\ and \mg2\ features (EW between $\sim$90 and $\sim$140 km s$^{-1}$; see their tab. 2 and fig. 2). With respect to the low-ionization transitions, the \al3\ spectral region is affected by fringing in the $8300 - 9000$ \AA\ range that artificially increases the flux level with respect to the reference quasar template, thus preventing a reliable estimate of the \al3\ absorption strength. The low-significance \mg2\ feature, composed by two troughs respectively at $v \sim 0$ (not sampled by the BI) and $\sim -2 \times 10^4$ km s$^{-1}$, is $\sim$10 to $\sim$20 times weaker than the \civ\ BAL when the AI or BI are used. Furthermore, any firm conclusion about the presence of \mg2\ BALs is prevented by the low-sensitivity gap between the bands $J$ and $H$. Therefore, we classify \hbox{\apm} as a high-ionization BAL QSO \hbox{(HiBAL)} due to the lack of unambiguous low-ionization absorption features.

\begin{figure*}[htbp]
\centering
\makebox[1\textwidth]{
	\subfigure[]{
		\includegraphics[width=0.5\linewidth]{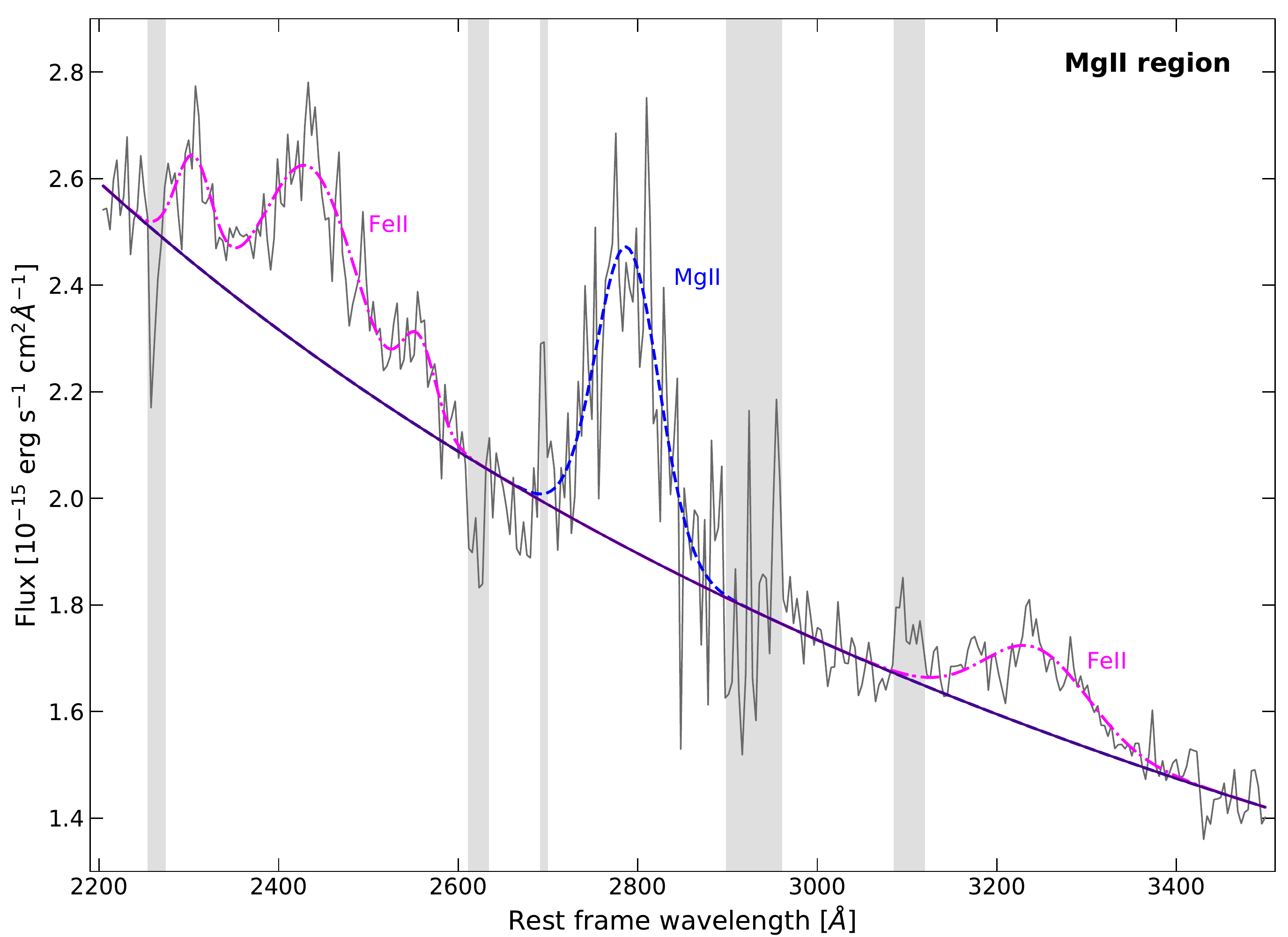}
		}
	\subfigure[]{
		\includegraphics[width=0.5\linewidth]{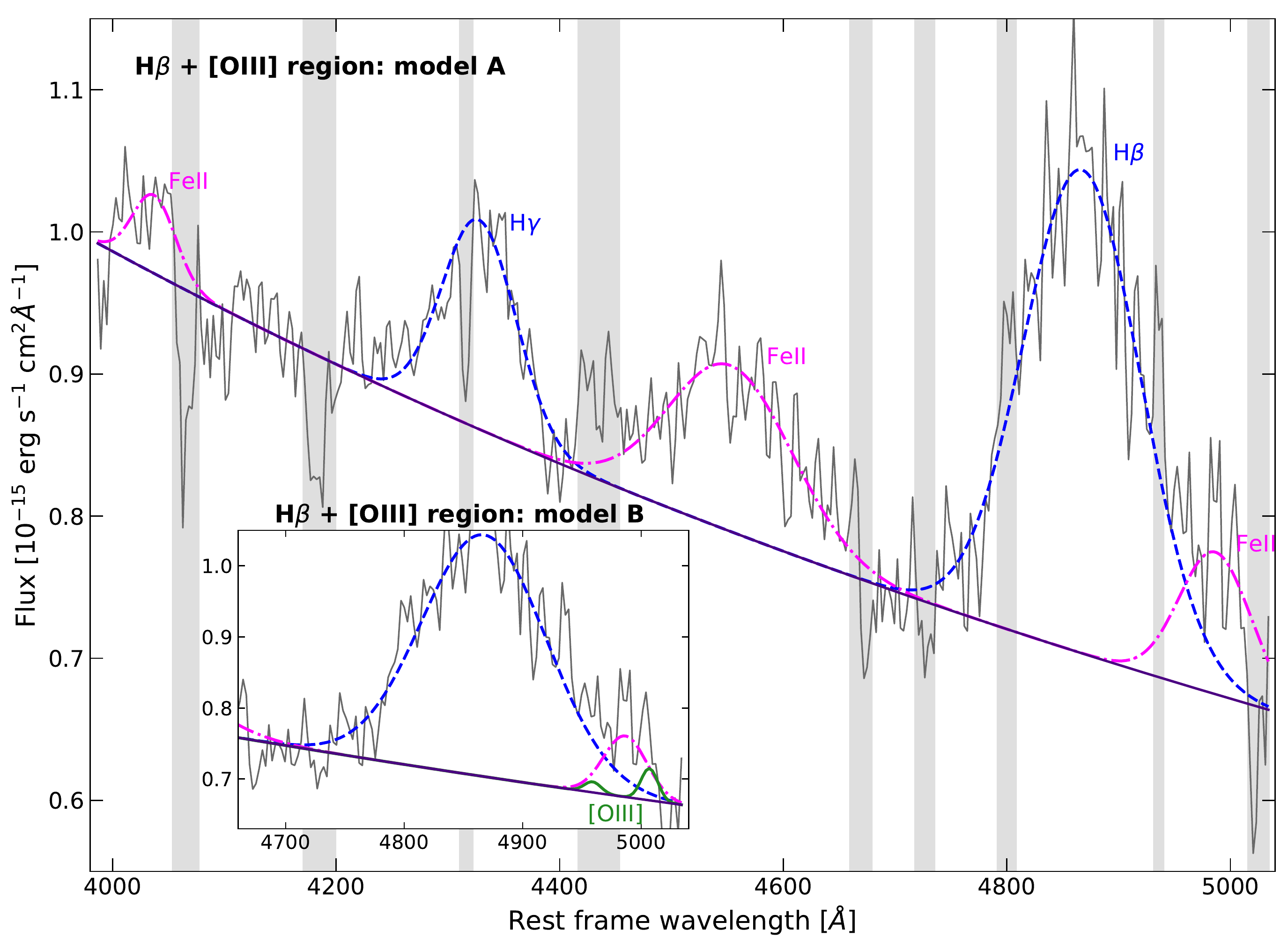}
		}     	
	}
\caption{NICS spectrum of \apm\ corresponding to the \mg2\ and \hbeta+\oiii\ spectral regions. Specifically, panel \textit{(a)} shows the best fit description of the rest-frame $\sim2200-3400$ \AA~band. Panel \textit{(b)} and its inset show the best-fit description of the rest-frame $\sim3990-5040$ \AA~band according to models A and B, respectively. In all panels, a purple solid line marks the power-law quasar continuum. Blue dashed curves refer to \mg2, H$\gamma$ and \hbeta\ emission lines, whereas the \fe2\ emission is plotted as magenta dot-dashed lines. In model B, the green curve refers to the \hbox{\oiii} emission. Grey bands indicate the regions excluded from the fit because of the presence of telluric features.}
\label{fig:fits}
\end{figure*}

HiBALs are the most common class of BAL QSOs \citep[$\sim$85\% of the BAL QSO population; e.g.,][]{Far07}. In particular, BALs associated with \civ\ represent the most common absorption troughs found in quasars, and are widely used to study the evolution of the BAL QSO population with cosmic time \citep{Hew03,Rei03,Kni08,Gib09,All11} as well as to characterize the {\itshape ensemble} absorption variability timescales \citep{Bar93,Lun07,Gib08,Gib10,Cap11,Cap12,Cap13,Fil14}. For instance, the time variability of the \apm\ \hbox{\civ} BAL has been studied in detail in \citet{Tre13} and \citet{Sat14,Sat16}, spanning a time interval of $\sim$19 yr in the observer frame (i.e. $\sim$3.9 yr in the rest frame). The discovery of significant low-ionization troughs would have been extremely interesting to e.g. unveil a possible transition between an obscured AGN phase and a normal quasar, as suggested by \citet{Far07,Far12} for the case of ultraluminous IR galaxies (ULIRGs) such as \apm\ \citep{Row00}. Nevertheless, the simultaneous presence of an UFO, a BAL and a molecular outflow in its spectrum makes \apm\ similar to the local quasar/ULIRG Mrk 231 \citep{Fer15}, configuring this object as one of the best potential targets to investigate multi-phase outflows at higher redshifts and extreme energetic regimes \citep[see e.g.][]{Cic18}.

\subsection{\mg2, H$\beta$ and \oiii\ emission lines}\label{weako3}

In order to study the properties of the \mg2, \hbeta\ and \oiii\ emission lines of \apm, we performed a spectral analysis of the two regions corresponding to the \hbox{\hbeta+\oiii} (rest-frame wavelength range $\sim$3990 -- 5040 \AA, which also includes the \hbox{H$\gamma$ $\lambda$4340 emission}) and \mg2\ (rest-frame wavelength range $\sim$2200 -- 3500 \AA). The analysis was done by using custom IDL processing scripts, based on the IDL package {\scriptsize MPFIT} \citep{Mark09}. The emission lines and the continuum emission were fitted together, by minimizing the $\chi^2$. Fig. \ref{fig:fits} shows a zoom out of such spectral regions, which are characterized by strong \hbox{\fe2} emission producing a complex pseudo-continuum close to \hbeta\ and \mg2.

As a first step, we tried to account for this \fe2-related emission in the rest-frame optical range by including the typical observational \fe2\ templates from \cite{BorGre92}, \cite{Ver04} and \cite{Tsu06} in the fits. We also considered the library of {\scriptsize CLOUDY}  \fe2\ synthetic spectral templates presented in \cite{Bis17} for hyper-luminous Type I quasars. In the fit, each template was convolved with a Gaussian whose width was free to vary, in order to account for the velocity dispersion of the gas. However, none of these templates was able to reproduce the \fe2\ spectral features observed in \hbox{\apm}. Furthermore, we also found that the observational \fe2\ template from \citet{Vest01} fails to reproduce the \fe2\ emission in the spectral region around \hbox{\mg2}. This failure might be partly due to the presence of telluric features in the near-IR spectra limiting the spectral windows used to anchor the fit. However, it is more likely that \hbox{\apm,} being an exceptional object, shows an intrinsic difference in the \fe2\ emission properties, as the relative intensities of the main \fe2\ emission blends do not match any template from the adopted library. Furthermore, we also tried to simultaneously fit two different \fe2\ templates with independent velocity dispersion, as done in Vietri et al. (2018) in case of hyper-luminous quasars similar to \apm\ with strong \fe2\ emission. However, this did not result in an improvement. Therefore, we fitted the most prominent \fe2\ features in the spectrum by means of multiple Gaussian components. We tried to limit the dependence of the resulting line parameters on the adopted model by using a minimum number of \fe2\ components, i.e. we checked that adding another Gaussian component did not lead to a significant improve of the $\chi^2$. However, a modest degeneracy between subtle \fe2\ emission and the other emission lines might still be present.

The spectral fitting procedure in the \hbeta+\oiii\ range is particularly challenging due to the combination of: (i) a limited spectral coverage redwards $5007$ \AA, (ii) the fact that these lines fall very close to the edge of the NICS wavelength coverage, which is affected by telluric absorption, and (iii) the lack of a clear emission feature at the wavelengths expected for the \hbox{\oiii} doublet. We thus tested two spectral models in order to infer the properties of any subtle \oiii\ emission. The first model (model A hereafter) only considers the presence of \fe2\ emission redwards the \hbeta\ emission, while the second model (model B) accounts for the presence of both \oiii\ and \fe2\ in this spectral region. Specifically, model A includes a Gaussian component with FWHM free to vary in order to account for the \fe2\ emission at $4861 < \lambda < 5000$ \AA~in the rest frame; the upper bound was set taking into account the NICS spectral resolution of $\sim$40 \AA\ at these wavelengths in order to not overlap with spectral regions involving possible \oiii\ emission at 5007 \AA.

Model B is similar to model A, but also includes two Gaussian components to fit the \oiii\ $\lambda\lambda$4959,5007 \AA\ doublet with a fixed ${\rm FWHM} = 1000$ km s$^{-1}$, which is a typical upper limit to the width of emission lines associated to the narrow-line region. The centroids of the \oiii\ doublet components were fixed to 5007 \AA\ and 4959 \AA\ in the rest frame, and the ratio of their normalizations was fixed to 1:3. Furthermore, both models include n total four Gaussian components to fit the \hbeta\ and H$\gamma$ broad emissions, the strong \fe2\ emission features centred at \hbox{$\sim$4050 \AA} and \hbox{$\sim$4550 \AA}, and a power law to parametrize the continuum emission. Apart from the \oiii\ doublet, the velocity offset between all Gaussian components are free to vary. Both models are shown in Fig. \ref{fig:fits}b.

\begin{table}[htbp]
\begin{center}
\resizebox{\columnwidth}{!}{
\begin{tabular}{lccr}
	\hline
	\hline
	 & & & \\
	Parameter & \multicolumn{2}{c}{\hbeta+\oiii\ region} &  Units\\
	\cline{2-3}
	\multicolumn{4}{l}{ }\\
						   	 & A & B &\\
	\multicolumn{4}{l}{ }\\
	\hline	
	\multicolumn{4}{l}{ }\\
	$\chi^{2}/N_{\rm d.o.f.}$  	 & $3863/2012$	& $3851/2011$ &  \\ 
	FWHM$\rm _{H\beta}$    	 & $6990 \pm 460$& $7360 \pm 430$ &  km s$^{-1}$\\
	$\rm \lambda_{H\beta}$ & $4866 \pm 7$ & $4868 \pm 7$ & \AA\\ 
	$F_{\rm [OIII]}(5007\mbox{ }\AA)$ & $-$ & $1.8 \pm 0.7$ & $10^{-15}$ erg s$^{-1}$ cm$^{-2}$ \\
	$\rm \lambda L_{\lambda}(5100\mbox{ }\AA)$ & $5.1 \pm 1.1$ & $5.1 \pm 1.2$ & $10^{47}$ erg s$^{-1}$ \\
	\multicolumn{4}{l}{ }\\
	\hbeta\ velocity shift & \multicolumn{2}{c}{$\lesssim 840$} & km s$^{-1}$\\
	\multicolumn{4}{l}{ }\\
	\hline
	 & & & \\
	Parameter & \multicolumn{2}{c}{\mg2\ region} & Units\\
	& & & \\
	\hline
	\multicolumn{4}{l}{ }\\
	$\chi^{2}/N_{\rm d.o.f.}$ & \multicolumn{2}{c}{$3186/2197$} & \\ 
	FWHM$\rm _{MgII}$ & \multicolumn{2}{c}{$9200_{-440}^{+610}$} & km s$^{-1}$\\
	$\rm \lambda_{MgII}$ & \multicolumn{2}{c}{$2789 \pm 4$} & \AA\\
	$\rm \lambda L_{\lambda}(3000\mbox{ }\AA)$ &\multicolumn{2}{c}{$7.9 \pm 1.9$} & $10^{47}$ erg s$^{-1}$\\
	& & & \\
	\mg2\ velocity shift & \multicolumn{2}{c}{$-1180 \pm 430$} & km s$^{-1}$\\
	
	\multicolumn{4}{l}{ }\\
	\hline
\end{tabular}
}
\end{center}
\caption{Spectral fit results derived for the \hbeta+\oiii\ and \mg2\ regions of \apm. }
\label{tab:fitres}
\end{table}

The rest-frame main spectral parameters, derived from the different models applied to the NICS data, are shown in Tab. \ref{tab:fitres}. The \hbeta\ emission is well reproduced by a broad Gaussian profile, with a $\rm FWHM_{H\beta} \sim 7000-7400$ km s$^{-1}$, that appears to be slightly redshifted ($\rm \lambda_{\rm H\beta} \sim 4868$ \AA, i.e. $\sim$410 km s$^{-1}$) with respect to the systemic redshift $z=3.911$ inferred from CO lines \citep{Dow99}; however, given the dispersion error of \hbox{$\sim$8 \AA} in the rest frame associated to the grism, the position of the \hbeta\ emission peak is still consistent with the assumed systemic redshift. The addition of the \oiii-related components in model B yields a decrease of $\Delta\chi^2 = 12$ for one additional free parameter (i.e. the \oiii\ $ \lambda$5007 \AA\ normalization) compared to model A, which represents a statistical improvement at 98.7\% confidence level according to an $F$-test. This suggests that a weak \oiii\ emission with \hbox{$F_{\rm [OIII]}({5007 \AA})$ = $(1.8\pm0.7) \times 10^{-15}$} erg s$^{-1}$ cm$^{-2}$, corresponding to an \hbox{\oiii-to-\hbeta} flux ratio $F_{\rm [OIII]}/F_{\rm H\beta} = 0.04$, can still be present in the optical spectrum of \apm. Such an \oiii\ weakness is consistent with the detection of strong Fe {\scriptsize II} emission, according to Eigenvector 1 \citep[e.g.,][]{BorGre92,She14}. However, we stress that any firm conclusion on the properties of \oiii\ emission in \hbox{\apm} is hampered by the low S/N and limited spectral coverage of the NICS data.

\begin{figure}[htbp]
\centering
\includegraphics[width=1\columnwidth]{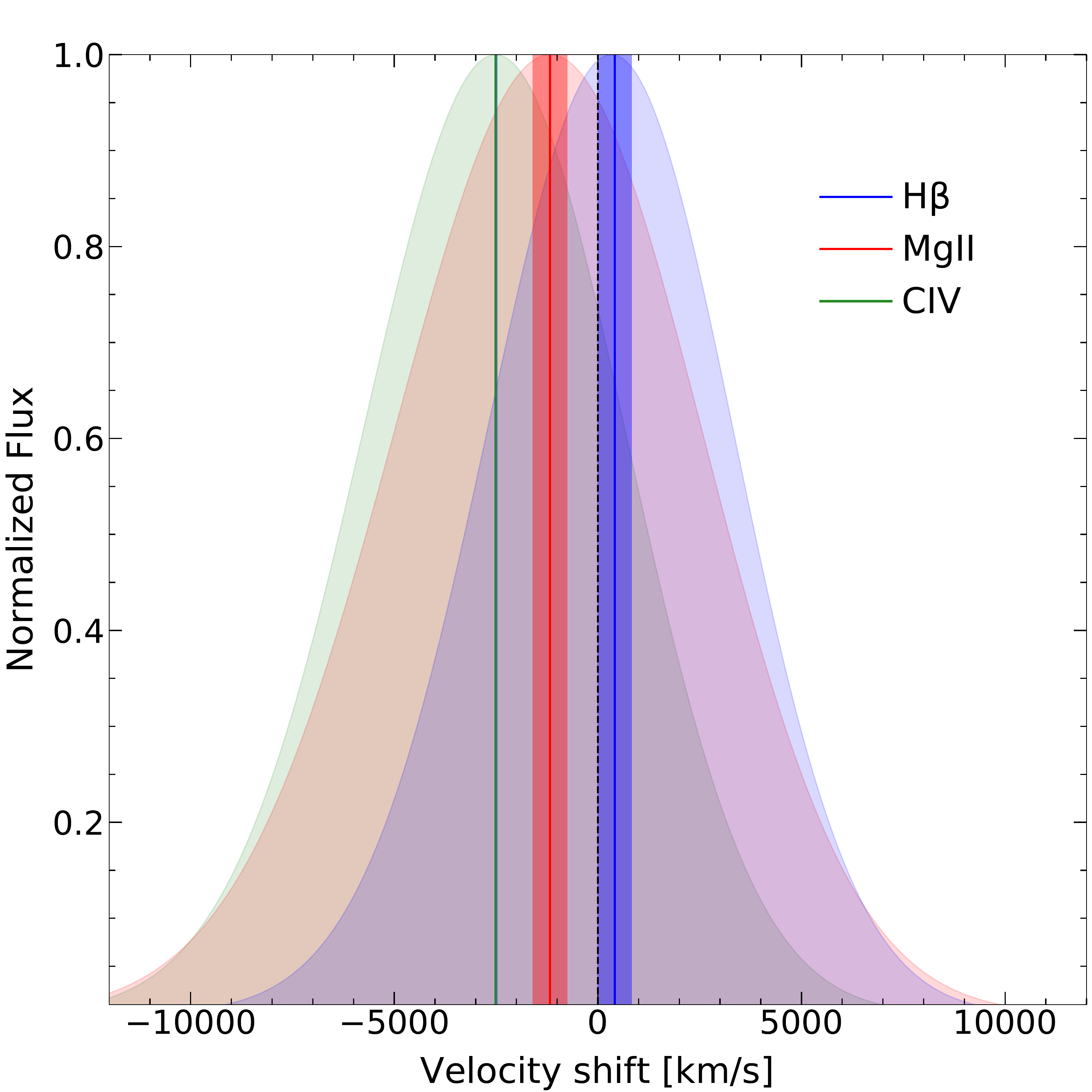}
\caption{Velocity shifts of the \hbeta\ ({\itshape blue}), \mg2\ ({\itshape red}) and \civ\ ({\itshape green}) lines with respect to $z = 3.911$. The centroids of the best fit Gaussian profiles resulting from our analysis (\hbeta\ and \mg2) and \citet[\hbox{\civ,} based on high-resolution {\itshape HST}/STIS data]{Sat16} are indicated by vertical lines with the corresponding uncertainties ({\itshape shaded vertical bands}).}
\label{fig:shift}
\end{figure}

As for the fit to the spectral data in the \mg2\ emission region, we used a model consisting of one Gaussian component that accounts for the \mg2\ line, four Gaussian components to fit the main \fe2\ emission features centered respectively at $\sim$2300 \AA, $\sim$2450 \AA, $\sim$2550 \AA\ and $\sim$3200 \AA\ in the rest frame, and a power law for the underlying continuum (see Fig. \ref{fig:fits}a). Such a fit yields a good description of the spectrum with an associated reduced $\chi^2 = 1.45$. The best-fit value for the FWHM of the \mg2\ broad emission line is $9200_{-440}^{+610}$ km s$^{-1}$. Remarkably, we found that the \mg2\ emission centroid is located at $\lambda_{\rm \mg2}\sim 2789$ \AA\, blueshifted by 11 \AA\ ($1180 \pm 430$ km s$^{-1}$) with respect to the expected value at the systemic redshift. This blueshift is a factor of $\sim$2 larger than the rest-frame dispersion error associated to the grism resolution in this spectral range, and is also $\sim$2 times smaller than the blueshift of $2500 \pm 40$ km s$^{-1}$ measured for the \civ\ emission \citep[and valid for other high-ionization emission lines such as N {\scriptsize V}, \si4\ and \c3; e.g.,][]{Irw98} from the $R$-band spectrum taken with the high-resolution spectrograph STIS on board of {\itshape HST} \citep{Lew02a}. We show this blueshift dependence on wavelength in Fig. \ref{fig:shift}, where the centroid shifts measured for \hbox{\civ,} \mg2\ and \hbeta\ broad emission lines are compared. This figure highlights that the \mg2\ emission line can be affected by a strong blueshift in the same way as high-ionization features. This can therefore bias the single-epoch SMBH mass estimate based on this low-ionization transition in case of highly-accreting AGN \citep[e.g.,][]{Mar13}. In particular, \mg2\ is extensively used to measure the SMBH mass of high-luminosity quasars at $z > 5$, due to the fact that \hbeta\ line is no longer observable in the $K$ band.

Previous studies found the \oiii\ weakness to be associated with broad blueshifted \civ\ emission in samples of Type 1 luminous quasars \citep[e.g.,][]{Net04}. In their study of the optical-to-UV spectra of WISE/SDSS selected hyper-luminous quasars \citep[WISSH; e.g.,][]{Bis17}, \citet{Vie18} have indeed found that $\sim$70\% of them exhibit very weak \oiii\ emission ($<$5 \AA) and largely blueshifted ($\approx 2000 - 8000$ \hbox{km s$^{-1}$}) \civ\ emission with rest-frame EWs $\lesssim$20 \AA. They interpreted these properties in terms of a steep UV-to-X-ray continuum in luminous quasars coupled to a face-on view of the continuum source. The former property leads to an efficient line-driving acceleration mechanism for broad-line region (BLR) winds \citep{Pro04,Wu09,Ris10,Ric11}, while the latter implies an observed small EW of the \oiii\ emission line \citep[see][]{Bio17}. The reconstruction of the \civ\ line profile on the \apm\ {\itshape HST}/STIS spectrum made by \citet{Sat16} allows us to compute a \hbox{\civ} rest-frame EW of $24 \pm 2$ \AA. This value is in agreement with \hbox{\civ} EWs commonly measured in other very luminous quasars with weak \oiii\ and large \hbox{\civ} blueshift.

\section{Single-epoch black hole mass estimates}\label{resconc}

The {\itshape TNG} observations of \apm\ allow us to provide the virial single-epoch black hole mass $M_{\rm BH}$ using \mg2\ and \hbeta\ emission lines. These lines are much more reliable proxies of the BLR dynamics than \civ, whose profile is potentially affected by non-virial motion of the emitting gas \citep{Bas05,SheLiu12,Vie18}. Furthermore, in the case of \hbox{\apm} the presence of strong BAL systems introduces additional uncertainty in the fit of the intrinsic \civ\ line profile. A direct measurement of $M_{\rm BH}$ based on \civ\ and \si4\ reverberation mapping \citep[RM; e.g.,][]{Pet97} was provided by \citet{Sat16}, who analyzed the {\itshape HST}/STIS spectrum. They were therefore able to accurately recover the \civ\ emission profile without the contamination of telluric features associated to the Fraunhofer A band on the red wing of the line, deriving a FWHM of $7480 \pm 70$ km s$^{-1}$ and, in turn, a $\log{M_{\rm BH}/{\rm M}_\odot} = 10.00^{+0.07}_{-0.05}$. The same RM-based value\footnote{Note that this result strictly holds only assuming a form factor \hbox{$f=5.5$} \citep{Onk04} for the high-ionization BLR. Other choices of the value of $f$ based on different calibrations of the BH mass-to-velocity dispersion relations \citep{Gra11,Par12,Gri13,She14} or on the study of BLR geometry \citep{Pan14,Pan14e} are possible.} of $M_{\rm BH}$ was also obtained for the \si4\ line, strengthening an estimate that could be affected by non-virial components as in the case of \civ.

In our calculations, we adopt the single-epoch relations derived by \citet{Bon14} for \mg2\ and \hbeta, using the FWHMs and reddening-corrected monochromatic luminosities listed in Tab. \ref{tab:fitres}. These relations can be expressed in the form:
\begingroup\makeatletter\def\f@size{8}\check@mathfonts
\begin{equation}\label{semass}
	\log{\left(
	\frac{M_{\rm BH}}{{\rm M}_\odot}
	\right)} = a + 0.5\log{\left[
	\frac{\lambda L_\lambda\left(
	{\rm 3000~\AA}/{\rm 5100~\AA}
	\right)}{10^{44}\mbox{ erg s}^{-1}}
	\right]}
	+ 2\log{\left(
	\frac{{\rm FWHM}_{\rm MgII/H\beta}}{1000\mbox{ km s}^{-1}}
	\right)},
\end{equation}
\endgroup
with the parameter $a = 6.6$ for \mg2\ and 6.7 for H$\beta$ respectively. In addition to the estimates of $M_{\rm BH}$ based on the \hbeta\ and \mg2\ from the {\itshape TNG} spectrum, we also derive an additional one based on \civ. We use the Gaussian fit to the \hbox{\civ} emission from the {\itshape HST}/STIS spectrum performed by \citet{Sat16} and adopt the single-epoch relation from \citet{Ves06} corrected according to the prescription on the line blueshift $\Delta v_{\rm CIV}$ for high-luminosity quasars \citep{Coa17}. This correction yields:
\begin{eqnarray}\label{sehl}
\log{\left(
\frac{M_{\rm BH}}{{\rm M}_\odot}
\right)} = 6.71 + 0.53 \log{\left[
\frac{\lambda L_\lambda\left(
{\rm 1350~\AA}
\right)}{10^{44}\mbox{ erg s}^{-1}}
\right]}\nonumber\\
 + 2 \log{\left(
\frac{{\rm FWHM}_{\rm CIV}}{1000\mbox{ km s}^{-1}}
\right)} - 2 \log{\left[
\alpha \left(
\frac{\Delta v_{\rm CIV}}{1000\mbox{ km s}^{-1}}
\right) + \beta
\right]},
\end{eqnarray}
with $\alpha \approx 0.4$ and $\beta \approx 0.6$.

Tab. \ref{tab:bhmass} shows the estimates of the $M_{\rm BH}$ of \apm\ based on the different transitions. Due to the fact that \hbox{\apm} is lensed by a foreground system which remains unobserved, at least two competing lens models have been proposed to explain the lack of a lens image. \citet{Ega00} proposed a naked-cusp configuration in which the magnification at UV-to-optical wavelengths can rise up to $\sim$100. \citet{Rie09} derived their lens model from the analysis of the CO emission region, finding a lower, almost achromatic magnification of $\sim$4. Accordingly, Tab. \ref{tab:bhmass} lists the SMBH mass estimates as a function of different values of the magnification parameter: $\mu = 1$ (i.e., no magnification), 4 and 100 respectively. The error associated to $\rm M_{\rm BH}$ includes in quadrature both the statistical uncertainties and intrinsic scatter in the single-epoch relations of $\sim$0.26 dex \citep{Ves06,She11,Bon14}. Even in the most conservative case of $\mu = 100$, the $M_{\rm BH}$ derived by considering the H$\beta$ emission line is $>$10$^{9}$ M$_\odot$, which indicates that \apm\ harbors a SMBH at the heaviest end of the $M_{\rm BH}$ distribution.

\begin{table}[htbp]
\begin{center}
\begin{tabular}{lccc}
		\hline
		\hline
		\multicolumn{4}{l}{} \\
	     \multicolumn{4}{c}{$\log{(M_{\rm BH}/{\rm M}_\odot)}$}\\
		\multicolumn{4}{l}{} \\
		 \hline
		Transition & $\mu = 1$ & $\mu = 4$ & $\mu = 100$\\
		\hline
		\multicolumn{4}{l}{} \\
		\hbeta\ & $10.29 \pm 0.32$ & $9.99 \pm 0.32$ & $9.29 \pm 0.32$\\
		\mg2\ & $10.45 \pm 0.34$ & $10.14 \pm 0.34$ & $9.45 \pm 0.34$\\
		\civ\ & $10.54 \pm 0.31$ & $10.22 \pm 0.31$ & $9.48 \pm 0.31$\\
		\multicolumn{4}{l}{} \\
		\hline
		\multicolumn{4}{l}{}\\
		 \multicolumn{4}{c}{$\lambda_{\rm Edd}$}\\
		\multicolumn{4}{l}{} \\
		\hline
		Transition & $\mu = 1$ & $\mu = 4$ & $\mu = 100$\\
		\hline
		\multicolumn{4}{l}{} \\
		\hbeta\ & $1.1\pm 0.8$ & $0.5\pm0.4$ & $0.11\pm0.08$\\
		\mg2\ & $0.8 \pm 0.6$ & $0.4\pm0.3$ & $0.08\pm0.05$\\
		\civ\ & $0.7 \pm 0.4$ & $0.4 \pm 0.2$ & $0.08 \pm 0.05$\\
		\multicolumn{4}{l}{}\\
		\hline
\end{tabular}
\end{center}
\caption{SMBH mass and Eddington ratio of \apm\ derived from \hbeta, \mg2\ and \civ\ broad emission lines, as a function of the magnification parameter $\mu$. The \civ-based estimates rely on the work by \citet{Sat16}. The Eddington ratio has been computed using a bolometric luminosity derived according to \citet{Run12e,Runn12}.}
\label{tab:bhmass}
\end{table}

We also report in Tab. \ref{tab:bhmass} the estimate of the Eddington ratios $\lambda_{\rm Edd} = L_{\rm bol}/L_{\rm Edd}$ based on these $M_{\rm BH}$ values, assuming a bolometric luminosity $L_{\rm bol} = 2.7 \times 10^{48}$ erg s$^{-1}$ before correcting for $\mu = 4$ and 100 (which correspond to a true bolometric luminosity of $6.8 \times 10^{47}$ erg s$^{-1}$ and $2.7 \times 10^{46}$ erg s$^{-1}$ respectively). We derive such a luminosity from the lensed monochromatic luminosity at 3000 \AA\ adopting the bolometric correction with non-zero intercept by \citet{Run12e,Runn12}:
\begin{equation}\label{eqn:runnoe}
L_{\rm bol}= 0.75 \left[10^{1.852} \cdot \lambda L_{\lambda}(3000\mbox{ }\AA)^{0.975}\right]
\end{equation}
We consider this estimate of $L_{\rm bol}$ as our fiducial value instead of deriving such a quantity from the fit of \apm\ spectral energy distribution (SED), since we note that the SED itself is affected by a wavelenght-dependent magnification factor resulting from the contribution of emitting regions of different size. In this way, we are also consistent with the UV-to-optical luminosities used to evaluate the single-epoch SMBH mass of \apm.

\citet{Don09} reported the existence of a strong correlation between the EW of \mg2\ and $\lambda_{\rm Edd}$, i.e. ${\rm EW_\mg2} \propto \lambda_{\rm Edd}^{-0.4}$. Such an effect is possibly due to a decrease in the covering factor of the \mg2\ BLR at increasing $\lambda_{\rm Edd}$ \citep[i.e., the number of \mg2\ emitting clouds decreases due to radiation-pressure blowing; see also][]{Fab06,Mar08b,Mar09}. We use this relation in order to get an independent estimate of $\lambda_{\rm Edd}$ for \apm. The rest-frame EW derived from the {\itshape TNG} spectrum is \hbox{${\rm EW_\mg2} = 27.1 \pm 1.5$} \AA, which corresponds to \hbox{$\lambda_{\rm Edd} = 0.36^{+0.51}_{-0.21}$} taking into account the scatter of 0.38 dex in $\lambda_{\rm Edd}$ in the \citet{Don09} data. We note that the confidence interval of $\lambda_{\rm Edd}$ obtained in this way is fully consistent with the values listed in Tab. \ref{tab:bhmass} for the case of $\mu = 4$, and only marginally consistent for the case of $\mu = 100$. This suggests that a moderate lens magnification $\mu < 100$ for \apm\ can be favored over a more extreme value.

We can derive an independent estimate of $\mu$ using the $\lambda_{\rm Edd}$ obtained from the \citet{Don09} relation. In fact, from \hbox{Eq. \ref{semass}} we derive that $\lambda_{\rm Edd} \propto \mu^{1 - b}$, and hence:
\begin{equation}\label{eqn:muedd}
\mu = \left[ \frac{\lambda_{\rm Edd}^{\rm (obs)}}{\lambda_{\rm Edd}^{\rm (true)}} \right]^{\frac{1}{1 - b}},
\end{equation}
where $b$ is the coefficient multiplying $\log{\left(\lambda L_\lambda\right)}$ in Eqs. \ref{semass} and \ref{sehl} ($0.5$ for \hbeta\ and \mg2\ from \citealt{Bon14}, and $0.53$ for \hbox{\civ} from \citealt{Ves06}). Therefore, a \hbox{$\lambda_{\rm Edd}^{\rm (true)} \sim 0.36$} corresponds to $4.1 \lesssim \mu \lesssim 9.3$ (and, in turn, a bolometric luminosity $2.9 \times 10^{47} \lesssim L_{\rm bol} \lesssim 6.7 \times 10^{47}$ erg s$^{-1}$ and a black hole mass $6.5 \times 10^9 \lesssim M_{\rm BH} \lesssim 1.5 \times 10^{10}$ M$_\odot$) using as $\lambda_{\rm Edd}^{\rm (obs)}$ the values listed in the first column of Tab. \ref{tab:bhmass} (i.e., the case of no magnification). Such an intermediate magnification factor is also in agreement with the upper limit of $\mu \lesssim 8.2$ derived by \citet{Sat16} from the RM. We also note that a lower limit on \apm\ black hole mass $M_{\rm BH} \gtrsim 4 \times 10^9$ M$_\odot$ has been derived by \citet{Hag17} through SED modeling. Using our fiducial luminosity and FWHM measurements for \mg2\ in Eq. \ref{semass} (see Tab. \ref{tab:fitres}), we find that this limit is respected for $\mu \lesssim 50$.

\section{Summary and conclusions}\label{disc}

In this work, we presented the quasi-simultaneous UV-to-optical spectrum of \apm\ taken at {\itshape TNG} with the instruments DOLoRes and NICS. The presence of a UFO, a BAL and a molecular outflow in this object is of great interest to explore the properties of multi-phase quasar winds at high redshifts and extreme luminosities with dedicated multi-wavelength observations, in order to probe the possible presence of ongoing AGN feedback. This spectrum covers the previously unobserved region between \c3\ and \oiii, thus providing important constraints on the BAL classification, the SMBH mass, the Eddington ratio and the magnification factor in this high-$z$ quasar. Our main results can be summarized as follows:

\begin{itemize}
	\item We tested the ``balnicity'' of \apm\ for high- and low-ionization transitions. We computed the most commonly used indexes (BI, BI$_0$ and AI) of absorption for \hbox{\si4}, \civ, \hbox{\al3} and \mg2, confirming the BAL only for \civ\ and hence supporting a HiBAL rather than a LoBAL classification for \apm.
	
	\item The near-infrared NICS spectrum allowed us, for the first time, to study the spectral regions corresponding to the \hbox{\hbeta+\oiii} and \mg2\ emission lines in \apm. The \hbeta\ line profile shows a FWHM of $\sim$7400 km s$^{-1}$ and  a centroid consistent with the CO-based systemic redshift $z = 3.911$. Conversely, the \mg2\ emission line \hbox{(${\rm FWHM} \sim 9200$ km s$^{-1}$)} is characterized by a blueshift of $\sim$1200 km \hbox{s$^{-1}$}, lower by a factor of $\sim$2 than the blueshift of $\sim$2500 \hbox{km s$^{-1}$} measured for the \civ\ emission line. This result is in agreement with previous works that find larger blueshifts in high-ionization transitions \citep{Ric02,Bas05}. 
		 
	\item We also investigated the presence of \oiii\ $\lambda\lambda$4959,5007 \AA~emission in a spectral region very close to the red edge of the NICS spectrum and characterized by strong \fe2\ emission. Our best-fit model includes a low-significance \oiii\ component with \hbox{$F_{\rm [OIII]}(5007\mbox{ }\AA)\sim (1.8 \pm 0.7) \times 10^{-15}$} \hbox{erg s$^{-1}$ cm$^{-2}$}. This indicates that the \oiii\ emission in \hbox{\apm} is intrinsically weak with \hbox{$F_{\rm [OIII]}/F_{\rm H\beta} = 0.04$}, consistent with the prediction of Eigenvector 1 \citep{BorGre92} of an anti-correlation between Fe {\scriptsize II} and \oiii\ emission. \hbox{\apm} therefore shares the properties of the \hbox{\oiii} and \civ\ emission lines observed in other sources lying  at the bright end of the AGN luminosity function \citep{Ric11,She14,Zuo15,Mar16,She16,Vie18}.

	\item We have been able to derive for the first time estimates of the $M_{\rm BH}$ in \apm\ based on the \hbeta\ and \mg2\ emission lines. These transitions have been found to provide a much more reliable measurement of $M_{\rm BH}$ in AGN than \hbox{\civ}. This is even more true in the case of \apm, whose \civ\ emission line is affected by BAL features. We find very large mass values ($\log{M_{\rm BH}/{\rm M}_\odot} \gtrsim 9.3$) for a magnification factor $\mu$ varying from 1 (i.e., no magnification) to 100 \citep{Ega00}. A value of $\mu=4$ \citep{Rie09} is compatible with the RM-based $M_{\rm BH}$ estimate of $\log{M_{\rm BH}/{\rm M}_\odot} \sim 10$ given in \cite{Sat16}. The ${\rm EW}_{\rm \mg2}$-based estimate of $\lambda_{\rm Edd}$ according to the ${\rm EW} - \lambda_{\rm Edd}$ relation by \citet{Don09} also suggests a moderate magnification factor $4.1 \lesssim \mu \lesssim 9.3$, corresponding to an intrinsic bolometric luminosity $2.9 \times 10^{47} \lesssim L_{\rm bol} \lesssim 6.7 \times 10^{47}$ erg s$^{-1}$ and a black hole mass $6.5 \times 10^9 \lesssim M_{\rm BH} \lesssim 1.5 \times 10^{10}$ M$_\odot$). This is in turn compatible with both the upper limit $\mu \lesssim 8.2$ found by \citet{Sat16} with RM, and the limit $\mu \lesssim 50$ derived from the single-epoch relations by adopting the minimal black hole mass $M_{\rm BH} = 4 \times 10^9$ M$_\odot$ inferred by \citet{Hag17} through \apm\ SED modeling.
\end{itemize}

Being taken 76 days apart (corresponding to $\sim$15 rest-frame days only), the UV and optical spectral sections of \hbox{\apm} probe the same AGN state. This allowed us to study its physical properties consistently during a state of relatively constant AGN emission. Indeed, this source has been found to vary in continuum, emission- and absorption-line intensity, with flux changes of up to $\sim$0.5 mag \citep{Tre13,Sat16}. Therefore, multi-band spectroscopic observations of \apm\ must be quasi-simultaneous in order to overcome the lack of a common reference for a compatible flux calibration between spectra taken at different epochs. The possibility to observe \apm\ with the forthcoming {\itshape James Webb Space Telescope} ({\itshape JWST}) is hence of extreme interest, since the spectro-photometric capabilities of its instruments, simultaneously covering the wavelength range \hbox{$\lambda\lambda 6000 - 2.8 \times 10^5$ \AA} \citep{Dor16,Lab16}, may allow for this source the detailed study of the stratified BLR dynamics around its central black hole and the characterization of the outflows associated to emission features from \si4\ to molecular lines.

Finally, we note that the estimates of $\mu$ and $\lambda_{\rm Edd}$ presented in this work are based on the comparison of the SE measurements of $M_{\rm BH}$ with the direct SMBH mass measurement by \citet{Sat16}, which relies, in turn, on the reverberation mapping of the high-ionization \civ\ and \si4\ lines that are affected by narrow (intrinsic or intervening) and broad absorption. Such absorption may bias the determination of $M_{\rm BH}$, and this bias is difficult to quantify. Future observations aimed at directly measuring \apm\ $M_{\rm BH}$ with novel techniques, such as BLR spectroastrometry, are therefore needed. In particular, the spectroastrometric technique presented in \citet{Ste15} may potentially allow to spatially resolve the kinematics of broad-line regions with single spectroscopic observations taken at high signal-to-noise ratio ($S/N \gtrsim 40$) in adaptive optics (AO) regime. Such capabilities are already at reach of current-generation telescopes such as the {\itshape Large Binocular Telescope} ({\itshape LBT}). Therefore, the AO infrared spectroscopy of \apm\ configures as a primary task to place better constraints on its SMBH mass and thus to obtain a more robust evaluation of lens magnification and accretion properties.

\newpage

\begin{acknowledgements}
We thank our anonymous referee for their helpful comments. We acknowledge R. Maiolino (Kavli Institute for Cosmology) and D. Trevese (``Sapienza'' University of Rome) for useful discussion. This research is based on observations made with the Italian {\itshape Telescopio Nazionale Galileo} ({\itshape TNG}) operated on the island of La Palma by the Fundaci{\'o}n Galileo Galilei of the INAF (Istituto Nazionale di Astrofisica) at the Spanish Observatorio del Roque de los Muchachos of the Instituto de Astrof{\'i}sica de Canarias. {\scriptsize IRAF} is distributed by the National Optical Astronomy Observatories, which are operated by the Association of the Universitiesfor Research in Astronomy, Inc., under cooperative agreement with the National Science Foundation. CC and CF acknowledge funding from the European Union's Horizon 2020 Research and Innovation Programme under the Marie Sklodowska-Curie grant agreement No. 664931. GV acknowledges funding support from the DFG Cluster of Excellence ``Origin and Structure of the Universe'' ({\ttfamily www.universe-cluster.de}). Reproduced with permission from Astronomy \& Astrophysics, \textcopyright~ESO.
\end{acknowledgements}

\bibliographystyle{aa}
\bibliography{apm_bib}

\end{document}